\documentclass[aps,twocolumn,pra,superscriptaddress,amsmath,showpacs,tightenlines]{revtex4}
\usepackage{epsfig,graphicx,times}
\usepackage{amstext}
\usepackage{amsmath}            
\usepackage{amssymb}            
\usepackage{graphicx}           
\usepackage{latexsym}
\usepackage{bm}
\begin{document}
\title{Antibunching photons in a cavity coupled to an optomechanical system}

\author{Xun-Wei Xu}
\affiliation{Institute of Microelectronics, Tsinghua University,
Beijing 100084, China}

\author{Yuan-Jie Li}
\affiliation{Department of Electronic Engineering, Tsinghua
University, Beijing 100084, China}

\date{\today}

\begin{abstract}
We study the photon statistics of a cavity linearly coupled to an
optomechanical system via second order correlation functions. Our
calculations show that the cavity can exhibit strong photon
antibunching even when optomechanical interaction in the
optomechanical system is weak. The cooperation between the weak
optomechanical interaction and the destructive interference between
different paths for two-photon excitation leads to the efficient
antibunching effect. Compared with the standard optomechanical
system, the coupling between a cavity and an optomechanical system
provides a method to relax the constraints to obtain single photon
by optomechanical interaction.
\end{abstract}

 \maketitle

\section{Introduction}

Photon antibunching is one of the evidences for the quantum nature of light, and the concept of photon blockade is introduced to explain the strong antibunching of transmitted photons~\cite{Imamoglu}. It is well known that such quantum effect can be observed in the strong nonlinear systems, such as, an optical cavity strongly coupled to a trapped atom~\cite{Birnbaum,Dayan}, a quantum dot strongly coupled to a photonic crystal resonator~\cite{Faraon}, and a superconducting qubit coupled to a microwave cavity in both resonant~\cite{Lang} and dispersive regime~\cite{Hoffman}. These systems provide a platform to realize non-classical photon states~\cite{Faraon}, which are of considerable interest for applications in quantum information processing and quantum cryptography~\cite{OBrien,Scarani}.

Optomechanics, a system that mechanical resonator acts as quantum system coupled to electromagnetic field via radiation pressure, provides a helpful toolbox for investigating the quantum effects in both optical and mechanical system (see, reviews~\cite{Kippenberg, Marquardt}). In recent years, many great experimental achievements have been obtained in this area, such as mechanical oscillator has been prepared almost in its ground state by sideband cooling, which paves the way for putting mechanical oscillators into quantum mechanics~\cite{Gigan, Arcizet, Schliesser, Wilson-Rae, MarquardtPRL, TeufelPRL, Thompson, Schliesser2008, Rocheleau, Groblacher2009, Park, Schliesser2009, Teufel475, Chan2009, Verhagen}. What's more, although the coupling constant between the optical and mechanical modes is weak in most standard optomechanical system, in the recent experiment, the strong optomechanical coupling has been obtained by driving the optomechanical system with an extra strong laser field~\cite{Groblacher}. In the meanwhile, the optical response of optomechanical systems to a signal field is modified by the driving field, leading to the effects such as normal-mode splitting~\cite{Groblacher, Teufel471} and electromagnetically induced transparency (EIT)~\cite{Teufel471, Agarwal, Weis, SafaviNaeini, Agarwal2012}.

The statistical properties of photons in optomechanical systems have been theoretically studied in Refs.~\cite{Rabl, Nunnenkamp}. These studies show that photon blockade effect can be observed in the optomechanical systems under the strong optomechanical coupling condition. Apart from blockade, photon can also induce multi-photon tunneling by the nonlinear interaction in optomechanical systems~\cite{xu-liu}. Moreover, it has been shown that the nonlinear interactions in two coupled optomechanical systems can be significantly enhanced for mechanical frequencies nearly resonant with the optical level splitting~\cite{Ludwig, Stannigel}. However, the photon blockade effects still only appear in the strong coupling regime, which is beyond the reach of most experiments in the single photon regime. Thus there is a question whether single-photon states can be generated using weak optomechanical interaction.

Recently, Liew and Savona have analyzed the photon statistics of two coupled nonlinear cavities, and found that
the photons can exhibit strong antibunching in such coupled systems with weak Kerr nonlinearity~\cite{Liew}. Later on,
such strong antibunching was attributed to the destructive quantum interference effect, and the authors further extended their theory to two coupled-cavities with a two-level quantum emitter embedded in one of cavities~\cite{Bamba}. They theoretically
demonstrated that perfect photon antibunching could be obtained even for single-atom cooperativity on the order of or smaller than unity. These studies~\cite{Liew,Bamba} have opened up a door towards nonlinear quantum optics at single-photon level using weak nonlinear coupling. More recently, a system with quantum dot coupled to a bimodal optical cavity has been proposed to achieve photon blockade in the weak coupling regime~\cite{Majumdar}.

Motivated by studies in Refs.~\cite{Liew,Bamba} and also recent progress in coupled-cavity and optomechanical systems, we now study the statistical properties of the photons in a cavity coupled to an optomechanical system, and show that the cavity can exhibit strong photon antibunching in the weak optomechanical interaction regime. The paper is organized as follows. In Sec.~II, the model Hamiltonian is introduced. In Sec.~III, the analytical expression of the second-order correlation function is obtained by the quantum Langevin equations under the semiclassical approximation, and we analyze the photon statistical properties of the cavity in Sec.~IV. In Sec.~V, we analyze the second-order correlation function further by numerical simulation via the master equation, and compare the results with those obtained under semiclassical approximation. Summary and conclusions are given in Sec.~VI.

\section{Model}

\begin{figure}
\includegraphics[bb=110 250 460 410, width=8 cm, clip]{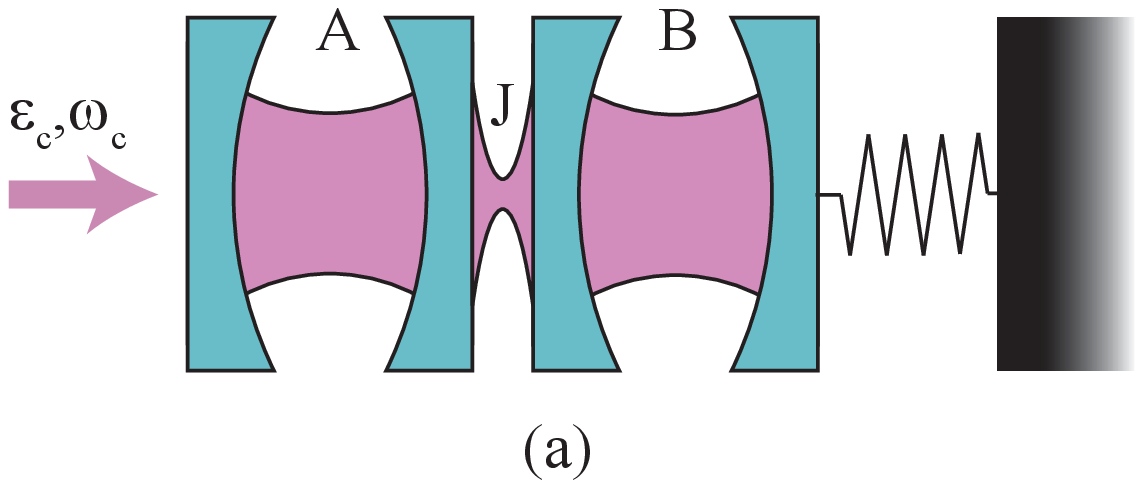}
\includegraphics[bb=25 166 588 567, width=8 cm, clip]{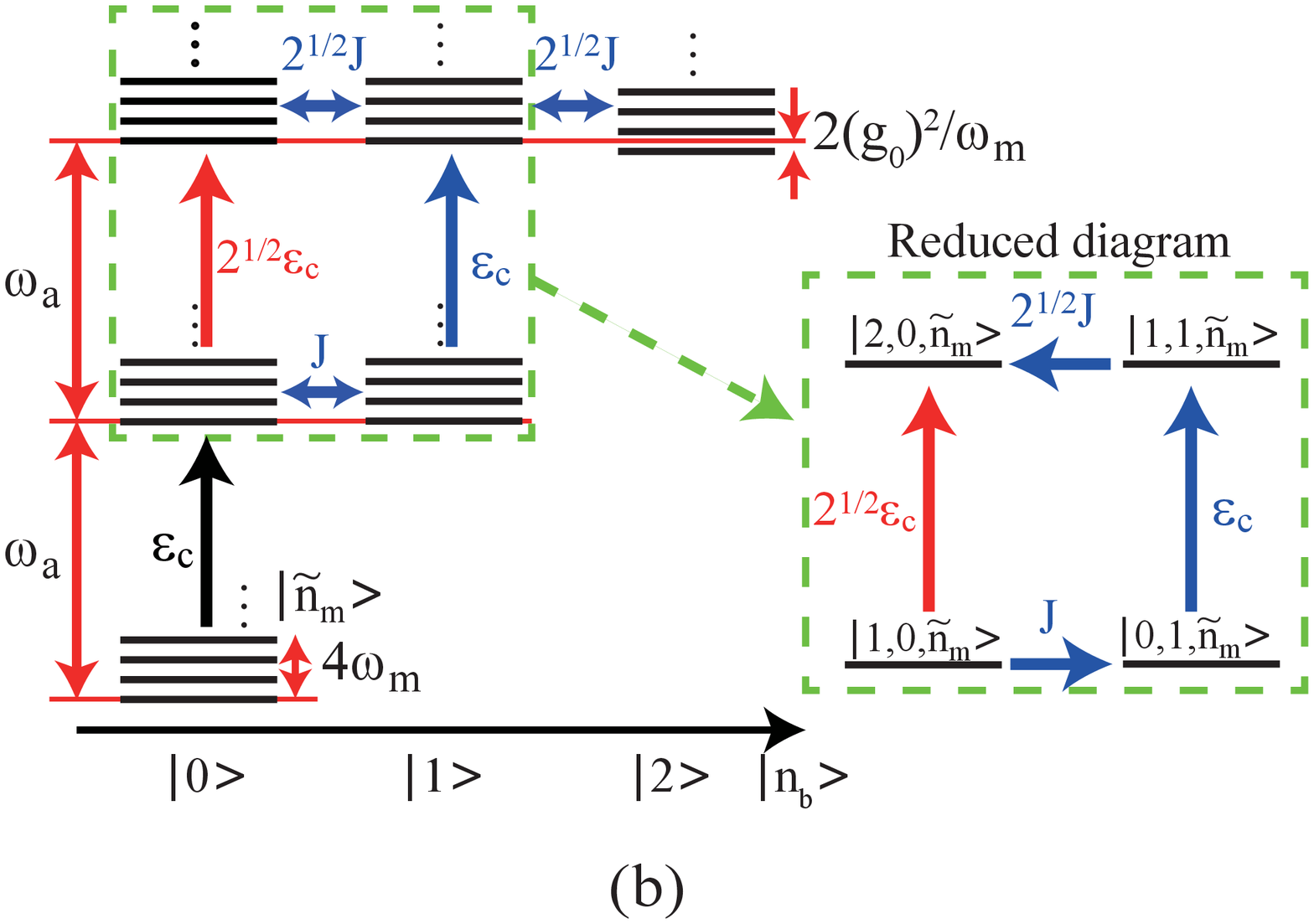}
\caption[]{(a) Schematic diagram for an optical cavity (cavity $A$, driven by a weak coherent laser field)
coupled to an optomechanical system (cavity $B$ with a movable right mirror). (b) Energy level diagram for the coupled system. Here, the short black lines denotes the energy levels $|n_{a},n_{b},\widetilde{n}_{m}\rangle$ for $\omega'_{b}=\omega _{a}$, and four levels are singled out as a reduced diagram (in the green dashed line box). $\varepsilon _{c}$ presents the coupling strength between the driving field and cavity field in cavity $A$. $J$ is the coupling constant between cavity $A$ and $B$.}\label{fig1}
\end{figure}

As schematically shown in Fig.~\ref{fig1}(a), the system consists of
two coupled cavities ($A$ and $B$) with the coupling constant
$J$. The cavity can be a transmission line resonator, a toroidal
microresonator, a cavity with two mirrors, or a defect cavity in
photonic crystal. Without loss of generality and for simplicity, we
will focus on the system of cavity with two mirrors. Cavity $A$ is driven by a
weak probe field with frequency $\omega_{c}$, and cavity $B$
consists of an oscillating mirror at one end, modeled as a quantum
mechanical harmonic oscillator. In other words, we study a coupled
system, which consists of a driven cavity and an optomechanical
system. The Hamiltonian of the whole system in the rotating wave
approximation is given as
\begin{eqnarray}\label{eq:1}
H &=&\hbar \omega _{a}a^{\dag }a+\hbar \omega _{b}b^{\dag }b+\hbar
\omega_{m}c^{\dag }c \nonumber\\
&&+\hbar J\left( a^{\dag }b+b^{\dag }a\right) +\hbar g_{0} b^{\dag
}b\left( c^{\dag }+c\right)  \nonumber\\
&&+i\hbar \varepsilon _{c}\left( a^{\dag }e^{-i\omega
_{c}t}-ae^{i\omega _{c}t}\right),
\end{eqnarray}%
where $a$ ($a^{\dagger}$) is the annihilation (creation) operator
for the light mode of the cavity $A$ with frequency $\omega_{a}$,
$b$ ($b^{\dagger }$) is the annihilation (creation) operator for the
light mode of the cavity $B$ with frequency $\omega_{b}$, and $c$
($c^{\dagger }$) is phonon annihilation (creation) operator of the
mechanically vibrational mode for the mirror with frequency
$\omega_{m}$. The parameter $g_{0}$ denotes the coupling strength
between the cavity $B$ and the oscillating mirror, and $\varepsilon
_{c}$ presents the coupling strength between the driving field and
cavity field inside the cavity $A$. As $\omega_{a}\approx \omega_{b}
\gg \omega_{m}, J$, we have dropped the rapidly varying terms
($ab$ and $a^{\dag}b^{\dag}$) corresponding to the rotating wave
approximation.

Our calculations (given in the following sections) show that cavity $A$ can exhibit strong photon antibunching effect even when optomechanical interaction in the cavity $B$ is weak. For the physical interpretation of the strong antibunching effect in the weak coupling condition, we are going to show the energy level diagram of the coupled system. It is convenient to change the Hamiltonian to a displaced oscillator representation $H_{\rm eff}=UHU^{\dag }$ by the unitary transformation $U=e^{-\frac{g_{0}}{\omega _{m}}b^{\dag }b\left( c^{\dag }-c\right) }$; then we obtain
\begin{eqnarray} \label{eq:2}
H_{\rm eff} & = &\hbar \omega _{a}a^{\dag }a+\hbar \omega'_{b}
b^{\dag }b- \hbar \frac{g_{0}^{2}}{\omega _{m}} b^{\dag }b^{\dag }bb+\hbar \omega _{m}c^{\dag }c  \notag \\
&&+\hbar J\left[ a^{\dag }b e^{\frac{g_{0}}{\omega _{m}}\left( c^{\dag
}-c\right) }+ab^{\dag }e^{-\frac{g_{0}}{\omega _{m}}\left( c^{\dag }-c\right)
}\right]  \notag \\
&&+i\hbar \varepsilon _{c}\left( a^{\dag }e^{-i\omega_{c}t}-ae^{i\omega _{c}t}\right),
\end{eqnarray}
where $\omega'_{b}=\omega_{b}-g_{0}^{2}/\omega _{m}$. In the limit $J, \varepsilon _{c} \rightarrow 0$, the Hamiltonian is diagonalized and the eigenvalues are
\begin{equation} \label{eq:3}
E_{n_{a},n_{b},n_{m}}=\hbar \omega _{a}n_{a}+\hbar \omega'_{b}n_{b}-\hbar\frac{g_{0}^{2}}{\omega_{m}}n_{b}(n_{b}-1)+\hbar \omega _{m} n_{m} ,
\end{equation}
corresponding to the eigenstates $|n_{a},n_{b},\widetilde{n}_{m}\rangle$, where $|n_{a},n_{b},\widetilde{n}_{m}\rangle \equiv U|n_{a},n_{b},n_{m}\rangle$ and $|n_{a},n_{b},n_{m}\rangle$ represents that there are $n_{a}$ ($n_{b}$) photons in cavity $A$ ($B$) and $n_{m}$ phonons in the mechanical resonator. The energy levels are shown by short black lines in Fig.\ref{fig1}(b) according to Eq.(\ref{eq:3}) by setting $\omega'_{b}=\omega _{a}$, and the terms for external driven ($\varepsilon_{c}$) and tunneling between the two cavities ($J$) are added and represented by lines with arrows in the diagram.

The optomechanical interaction in cavity $B$ and the quantum
interference effect between the two cavities (cavity $A$ and cavity
$B$) are responsible for the photon antibunching
effect~\cite{Bamba}. As shown in the reduced diagram in Fig.\ref{fig1}(b), the interference is between two paths for
two-photon excitation in cavity $A$: (i) the direct excitation from
one photon to two photons in the cavity $A$; and (ii) one photon
tunneling from cavity $A$ to cavity $B$, then exciting another
photon in cavity $A$, and finally the photon inside cavity $B$
tunneling back to cavity $A$. The destructive interference
between the two paths reduces the probability of two-photon excitation
in cavity $A$.

In order to analyze this phenomenon more precisely, the second-order correlation function is calculated by the quantum Langevin equations under the semiclassical approximation and by numerical simulation via the master equation in the following sections. To remove the time-dependent factor, let us transform the Hamiltonian in Eq.~(\ref{eq:1}) into the rotating reference frame
through a unitary operator $R(t)=\exp[-i\omega_{c}t(a^{\dag }a+b^{\dag }b)]$, and thus Eq.~(\ref{eq:1}) becomes
\begin{eqnarray}\label{eq:4}
\widetilde{H} &=&\hbar \Delta _{a}a^{\dag }a+\hbar \Delta_{b} b^{\dag }b+\hbar \omega_{m}c^{\dag }c \nonumber \\ &&+\hbar
J\left( a^{\dag}b+b^{\dag}a\right) +\hbar g_{0} b^{\dag
}b\left(c^{\dag }+c\right) \nonumber \\ &&+i\hbar
\varepsilon_{c}\left(a^{\dag}-a\right),
\end{eqnarray}%
where $\Delta _{a}=\omega _{a}-\omega _{c}$ and $\Delta _{b}=\omega
_{b}-\omega_{c}$ are the detunings of the frequencies of cavity
fields from that of the driving field.

\section{Langevin equations and second order correlation functions}

The dynamics of the cavity fields and mechanical oscillator can be
described by quantum Langevin equations. By considering the
dissipation and fluctuation of the light fields and
mechanical mode, we can write out a set of nonlinear quantum
Langevin equations as follows
\begin{eqnarray}
\frac{d}{dt}a &=&-\left( \frac{\kappa_{a}}{2}+i\Delta _{a}\right)
a-iJb+\varepsilon _{c}+\sqrt{\kappa _{a}}a_{\rm in}, \label{eq:5}\\
\frac{d}{dt}b &=&-\left[ \frac{\kappa _{b}}{2}+i\left( \Delta
_{b}+g_{b}q\right) \right] b-iJa+\sqrt{\kappa _{b}}b_{\rm in}, \label{eq:6}\\
\frac{d}{dt}q &=&\omega _{m}p, \label{eq:7}\\
\frac{d}{dt}p &=&-\omega _{m}q-g_{b}b^{\dag }b-\frac{\gamma
_{m}}{2}p+\xi,\label{eq:8}
\end{eqnarray}
where $\kappa _{a}$, $\kappa _{b}$ and $\gamma _{m}$ are the damping
rates of cavity $A$, cavity $B$, and the moving mirror,
respectively. $q=\left( c+c^{\dag }\right)/\sqrt{2} $,
$p=\left(c-c^{\dag }\right)/(i\sqrt{2})$, and
$g_{b}=\sqrt{2}g_{0}$. $\xi$ is a Brownian stochastic force with zero
mean value, i.e. $\langle \xi \left( t\right)\rangle=0$ , which
comes from the coupling of the oscillating mechanical resonator to
its thermal environment and satisfies
correlation~\cite{Ford1965,Ford1988,Giovannetti,Genes}
\begin{equation}%
\left\langle \xi \left( t\right) \xi \left( t^{\prime }\right)
\right\rangle =\frac{\gamma _{m}}{2\omega _{m}}\int \frac{d\omega
}{2\pi }\omega e^{-i\omega \left( t-t^{\prime }\right) }\left[
1+\coth \left( \frac{\hbar \omega }{2k_{B}T}\right) \right],\label{eq:9}
\end{equation}%
where $k_{B}$ is the Boltzmann constant and $T$ is the effective
temperature of the environment of the mechanical resonator. $a_{\rm
in}$ and $b_{\rm in}$ represent the vacuum radiation noises input to
the cavity $A$ and $B$ with $\langle a_{\rm in}\left(
t\right)\rangle=\langle b_{\rm in}\left( t\right)\rangle=0$, and
they obey the following correlation functions~\cite{Gardiner}
\begin{eqnarray}%
\left\langle a_{\rm in}^{\dag }\left( t\right) a_{\rm in}\left(
t^{\prime
}\right)\right\rangle  &=& 0, \label{eq:10} \\
\left\langle a_{\rm in}\left( t\right) a_{\rm in}^{\dag }\left(
t^{\prime }\right)
\right\rangle  &=& \delta \left( t-t^{\prime }\right), \label{eq:11}\\
\left\langle b_{\rm in}^{\dag }\left( t\right) b_{\rm in}\left(
t^{\prime
}\right)\right\rangle  &=&0, \label{eq:12} \\
\left\langle b_{\rm in}\left( t\right) b_{\rm in}^{\dag }\left(
t^{\prime }\right) \right\rangle  &=& \delta \left( t-t^{\prime
}\right).\label{eq:13}
\end{eqnarray}%
Here, we have assumed that the whole system is in a low temperature
environment, and therefore the equilibrium mean thermal photon numbers
in two cavities at optical frequencies have been neglected.

The dynamic of the system is determined by the small
fluctuations when the system reaches the steady-state. Thus, let us
now apply semiclassical approximation to solve the steady-state with
small quantum fluctuations. That is, we assume $a=\alpha_{0} +\delta
a$, $b=\beta_{0} +\delta b$, $q=q_{0}+\delta q$, here $\alpha_{0}$,
$\beta_{0}$ and $q_{0}$ are the mean values of the cavity fields and
mechanical mode when the system reaches the steady-state, and operators
$\delta a$, $\delta b$ and $\delta q$ describe the small
fluctuations around steady-state with zero mean value, $\langle
\delta a\rangle =0$, $\langle\delta b\rangle =0$ and $\langle\delta
q\rangle =0$. The steady-state values satisfy the following
equations
\begin{eqnarray}%
&&\left( \frac{\kappa _{a}}{2}+i\Delta _{a}\right) \alpha
_{0}+iJ\beta
_{0}=\varepsilon _{c}, \label{eq:14}\\
&&\left[ \frac{\kappa _{b}}{2}+i\left( \Delta _{b}+g_{b}q_{0}\right)
\right]
\beta _{0}+iJ\alpha _{0}=0,\label{eq:15} \\
&&\omega _{m}q_{0}=-g_{b}\left\vert \beta _{0}\right\vert
^{2}.\label{eq:16}
\end{eqnarray}%
Here, we have used the factorization assumption, e.g., $\langle
qb\rangle =\langle q \rangle \langle b \rangle$. The dynamics of
small fluctuations around steady-state can be obtained by
linearizing Eqs.~(\ref{eq:5}-\ref{eq:8}) as
\begin{eqnarray}
\frac{d}{dt}\delta a &=&-\left( \frac{\kappa _{a}}{2}+i\Delta _{a}\right)
\delta a-iJ\delta b+\sqrt{\kappa _{a}}a_{\rm in}, \label{eq:17}\\
\frac{d}{dt}\delta b &=&-\left[ \frac{\kappa _{b}}{2}+i\left( \Delta
_{b}+g_{b}q_{0}\right) \right] \delta b-ig_{b}\beta _{0}\delta q \nonumber\\
&&-iJ\delta a+\sqrt{\kappa _{b}}b_{\rm in}, \label{eq:18}\\
\frac{d}{dt}\delta q &=&\omega _{m}\delta p, \label{eq:19}\\
\frac{d}{dt}\delta p &=&-\omega _{m}\delta q-g_{b}\left( \beta
_{0}^{\ast }\delta b+\beta _{0}\delta b^{\dag }\right) -\frac{\gamma
_{m}}{2}\delta p+\xi, \label{eq:20}
\end{eqnarray}
here, the high order terms of small fluctuations, e.g., $\delta
q\delta b$, have been neglected. The system is stable only if all
the eigenvalues of the coefficient matrix of the above differential
equations have negative real parts, and the stability condition can
be given explicitly by using the Routh-Hurwitz
criterion~\cite{DeJesus}. However, it is too cumbersome to be given
here. All the parameters we will use satisfy the stability
condition, and it is easy to fulfill for the driving field in our
system is weak.

By applying the Fourier transform and solving dynamical equations in
the frequency domain, we obtain
\begin{eqnarray} \label{eq:21}
\delta a (\omega) &=& E\left( \omega \right) a_{\rm in}\left( \omega
\right) +F\left( \omega \right) a_{\rm in}^{\dag }\left( \omega \right) \nonumber\\
&&+G\left( \omega \right) b_{\rm in}\left( \omega \right)+H\left( \omega
\right) b_{\rm in}^{\dag }\left( \omega \right) \nonumber\\
&&+Q\left( \omega \right) \xi \left( \omega \right),
\end{eqnarray}
where
\begin{eqnarray}
E\left( \omega \right)  &=&\sqrt{\kappa _{a}}\frac{A_{11}(\omega)}{D(\omega)}, \label{eq:22}\\ %
F\left( \omega \right)  &=&-\sqrt{\kappa _{a}}\frac{A_{22}(\omega)}{D(\omega)}, \label{eq:23}\\ %
G\left( \omega \right)  &=&\sqrt{\kappa _{b}}\frac{A_{33}(\omega)}{D(\omega)}, \label{eq:24}\\ %
H\left( \omega \right)  &=&-\sqrt{\kappa _{b}}\frac{A_{44}(\omega)}{D(\omega)}, \label{eq:25}\\ %
Q\left( \omega \right)  &=&-i\frac{g_{b}\chi \left( \omega \right) }{\omega _{m}D(\omega)}%
                           \left[\beta _{0}A_{33}(\omega)+\beta _{0}^{\ast}A_{44}(\omega)\right], \label{eq:26}%
\end{eqnarray}%
and
\begin{eqnarray}
A_{11}(\omega) &=&\left[ \frac{\kappa _{a}}{2}-i\left( \Delta _{a}+\omega \right) %
\right] \left[ \left( \frac{\kappa _{b}}{2}-i\omega \right)^{2}+\Delta
_{b}^{\prime 2}\right]  \nonumber \\
&&-\left[ \frac{\kappa _{a}}{2}-i\left( \Delta _{a}+\omega \right)
\right] g_{b}^{4}\left\vert \beta _{0}\right\vert ^{4}\left(
\frac{\chi \left(
\omega \right) }{\omega _{m}}\right) ^{2} \nonumber \\
&&+J^{2}\left[ \frac{\kappa _{b}}{2}+i\left( \Delta _{b}^{\prime
}-\omega \right) \right],  \label{eq:27}\\
A_{22}(\omega)&=&-iJ^{2}g_{b}^{2}\left( \beta _{0}\right)
^{2}\frac{\chi \left(
\omega \right) }{\omega _{m}}, \label{eq:28}\\
A_{33}(\omega) &=&-iJ\left[ \frac{\kappa _{a}}{2}-i\left( \Delta
_{a}+\omega \right) \right] \left[ \frac{\kappa _{b}}{2}-i\left(
\Delta _{b}^{\prime
}+\omega \right) \right] \nonumber\\
&&-iJ^{3}, \label{eq:29}\\
A_{44}(\omega) &=&-Jg_{b}^{2}\left( \beta _{0}\right)
^{2}\frac{\chi \left( \omega \right) }{\omega _{m}}\left[
\frac{\kappa _{a}}{2}-i\left( \Delta
_{a}+\omega \right) \right],  \label{eq:30}\\
D(\omega)&=&\left[ \frac{\kappa _{a}}{2}+i\left( \Delta _{a}-\omega
\right) \right] A_{11}(\omega)+iJA_{33}(\omega). \label{eq:31}
\end{eqnarray}%
Here, we introduce $ \Delta _{b}^{\prime }=\Delta
_{b}+g_{b}q_{0}-g_{b}^{2}\left\vert \beta _{0}\right\vert
^{2}\frac{\chi \left( \omega \right) }{\omega _{m}} $, and the
dynamical response function of the mirror~\cite{Mancini94}
\begin{equation} \label{eq:32}
\chi \left( \omega \right) =\frac{\omega _{m}^{2}}{\left( \omega
_{m}^{2}-\omega ^{2}-i\omega \gamma _{m}/2\right) }
\end{equation}%
with $\chi ^{\ast}\left( \omega \right)=\chi \left( -\omega
\right)$.

The second order correlation functions that can be measured outside the cavity have the structure of time-antiordered product followed by a time-ordered product, which are called multitime ordered correlation functions~\cite{Walls, Gardiner1985}, thus
\begin{equation}\label{eq:33}
g_{aa}^{\left( 2\right) }\left( \tau \right) =\frac{\left\langle \widetilde{T%
}\left[ a^{\dag }\left( t\right) a^{\dag }\left( t+\tau \right) \right] T%
\left[ a\left( t+\tau \right) a\left( t\right) \right] \right\rangle }{%
\left\langle a^{\dag }\left( t\right) a\left( t\right) \right\rangle
\left\langle a^{\dag }\left( t+\tau \right) a\left( t+\tau \right)
\right\rangle} ,
\end{equation}%
where $\widetilde{T}$ is the time-antiordered product and $T$ the time-ordered product.
By taking $a=\alpha _{0}+\delta a$, the second-order correlation function of the light field in the cavity $A$, $g_{aa}^{\left( 2\right) }\left(
\tau \right)$, can be given as
\begin{equation} \label{eq:34}
g_{aa}^{\left( 2\right) }\left( \tau \right) =G_{1}\left( \tau \right)
+G_{2}\left( \tau \right),
\end{equation}%
where
\begin{eqnarray}
G_{1}\left( \tau \right)  &=&\frac{\left\vert \alpha \right\vert
^{4}+2\left\vert \alpha \right\vert ^{2}\left\langle \delta a^{\dag }\left(
t\right) \delta a\left( t\right) \right\rangle }{\left( \left\vert \alpha
\right\vert ^{2}+\left\langle \delta a^{\dag }\left( t\right) \delta a\left(
t\right) \right\rangle \right) ^{2}}  \notag \\
&&+\frac{2\left\vert \alpha \right\vert ^{2}{\rm Re}\left[ \left\langle
\delta a^{\dag }\left( t\right) \delta a\left( t+\tau \right) \right\rangle %
\right] }{\left( \left\vert \alpha \right\vert ^{2}+\left\langle \delta
a^{\dag }\left( t\right) \delta a\left( t\right) \right\rangle \right) ^{2}}
\notag \\
&&+\frac{2 {\rm Re}\left[ \left( \alpha ^{\ast }\right) ^{2}\left\langle T%
\left[ \delta a\left( t+\tau \right) \delta a\left( t\right) \right]
\right\rangle \right] }{\left( \left\vert \alpha \right\vert
^{2}+\left\langle \delta a^{\dag }\left( t\right) \delta a\left( t\right)
\right\rangle \right) ^{2}}, \\  \label{eq:35}
G_{2}\left( \tau \right)  &=&\frac{\left\langle \delta a^{\dag }\left(
t\right) \delta a\left( t\right) \right\rangle ^{2}+\left\vert \left\langle
\delta a^{\dag }\left( t\right) \delta a\left( t+\tau \right) \right\rangle
\right\vert ^{2}}{\left( \left\vert \alpha \right\vert ^{2}+\left\langle
\delta a^{\dag }\left( t\right) \delta a\left( t\right) \right\rangle
\right) ^{2}}  \notag \\
&&+\frac{\left\vert \left\langle T\left[ \delta a\left( t+\tau \right)
\delta a\left( t\right) \right] \right\rangle \right\vert ^{2}}{\left(
\left\vert \alpha \right\vert ^{2}+\left\langle \delta a^{\dag }\left(
t\right) \delta a\left( t\right) \right\rangle \right) ^{2}}.  \label{eq:36}
\end{eqnarray}%
$G_{2}\left( \tau \right)$ comes from the four-operator correlation and is obtained by
the properties of the Gaussian process~\cite{Ford1965,Ford1988}. Using
the expression of $\delta a(t)$, Eqs.~(\ref{eq:21}), and the
correlations Eqs.~(\ref{eq:9}-\ref{eq:13}), the correlation of
$\delta a(t)$ and $\delta a^{\dag}(t)$ are given by
\begin{equation}
\left\langle \delta a^{\dag }\left( t\right) \delta a\left( t+\tau \right)
\right\rangle =\frac{1}{2\pi }\int_{-\infty }^{+\infty }X_{a^{\dag }a}\left(
\omega \right) e^{i\omega \tau }d\omega , \label{eq:37}
\end{equation}%
\begin{equation}
\left\langle T\left[ \delta a\left( t+\tau \right) \delta a\left( t\right) %
\right] \right\rangle =\frac{1}{2\pi }\int_{-\infty }^{+\infty }X_{aa}\left(
\omega \right) e^{-i\omega \left\vert \tau \right\vert }d\omega , \label{eq:38}
\end{equation}%
where
\begin{eqnarray}
X_{a^{\dag }a}\left( \omega \right)  &=&\left\vert Q\left( -\omega \right)
\right\vert ^{2}\frac{\gamma _{m}}{2\omega _{m}}\omega \left[ 1+\coth \left(
\frac{\hbar \omega }{2k_{B}T}\right) \right]   \notag \\
&&+\left\vert F\left( -\omega \right) \right\vert ^{2}+\left\vert H\left(
-\omega \right) \right\vert ^{2} , \label{eq:39}
\end{eqnarray}%
\begin{eqnarray}
X_{aa}\left( \omega \right)  &=&Q\left( \omega \right) Q\left( -\omega
\right) \frac{\gamma _{m}}{2\omega _{m}}\omega \left[ 1+\coth \left( \frac{%
\hbar \omega }{2k_{B}T}\right) \right]   \notag \\
&&+E\left( \omega \right) F\left( -\omega \right) +G\left( \omega \right)
H\left( -\omega \right) . \label{eq:40}
\end{eqnarray}

\section{Statistical properties of the field in cavity $A$}

\begin{figure}
\includegraphics[bb=5 1 393 275, width=4 cm, clip]{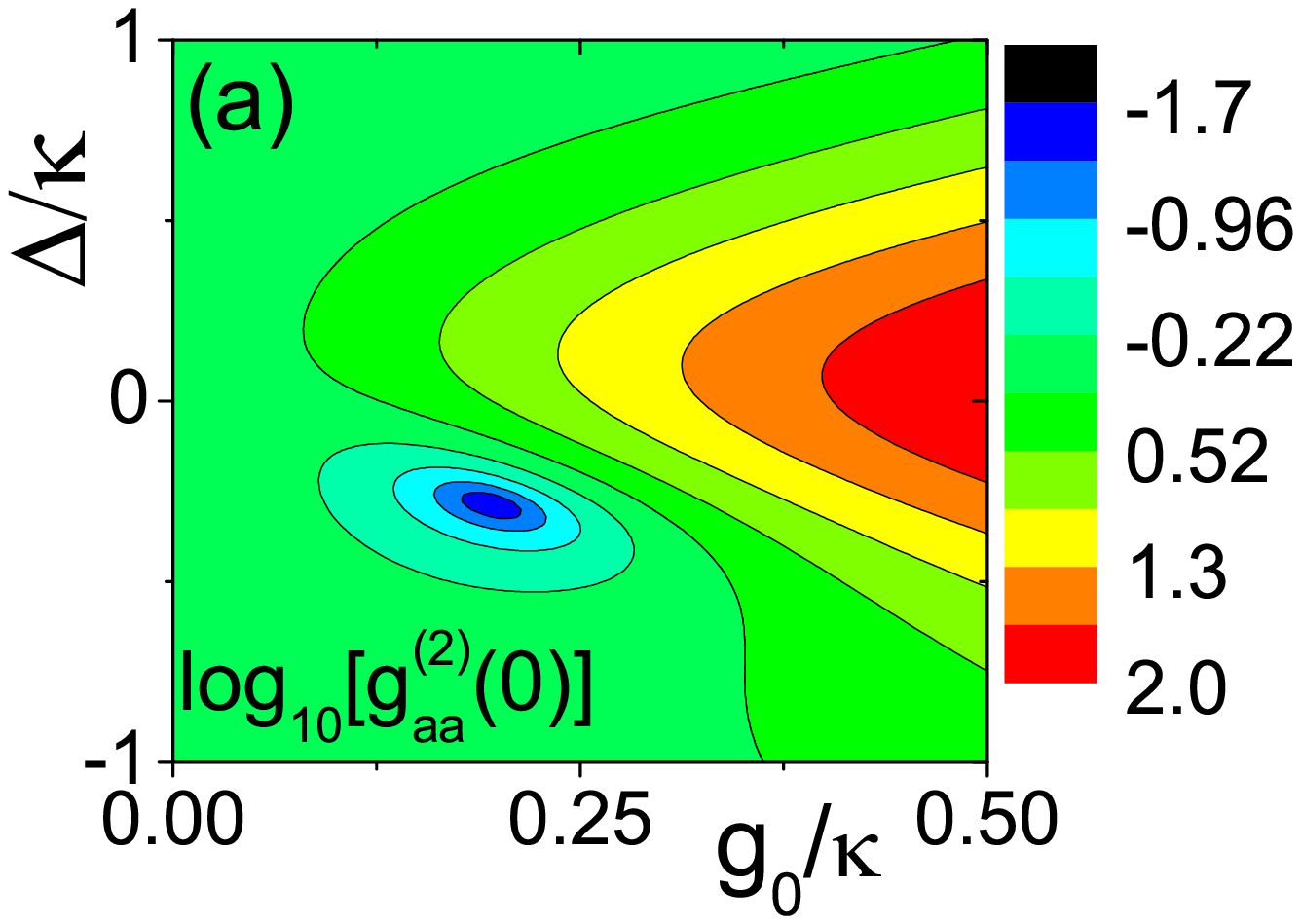}
\includegraphics[bb=1 8 393 275, width=4 cm, clip]{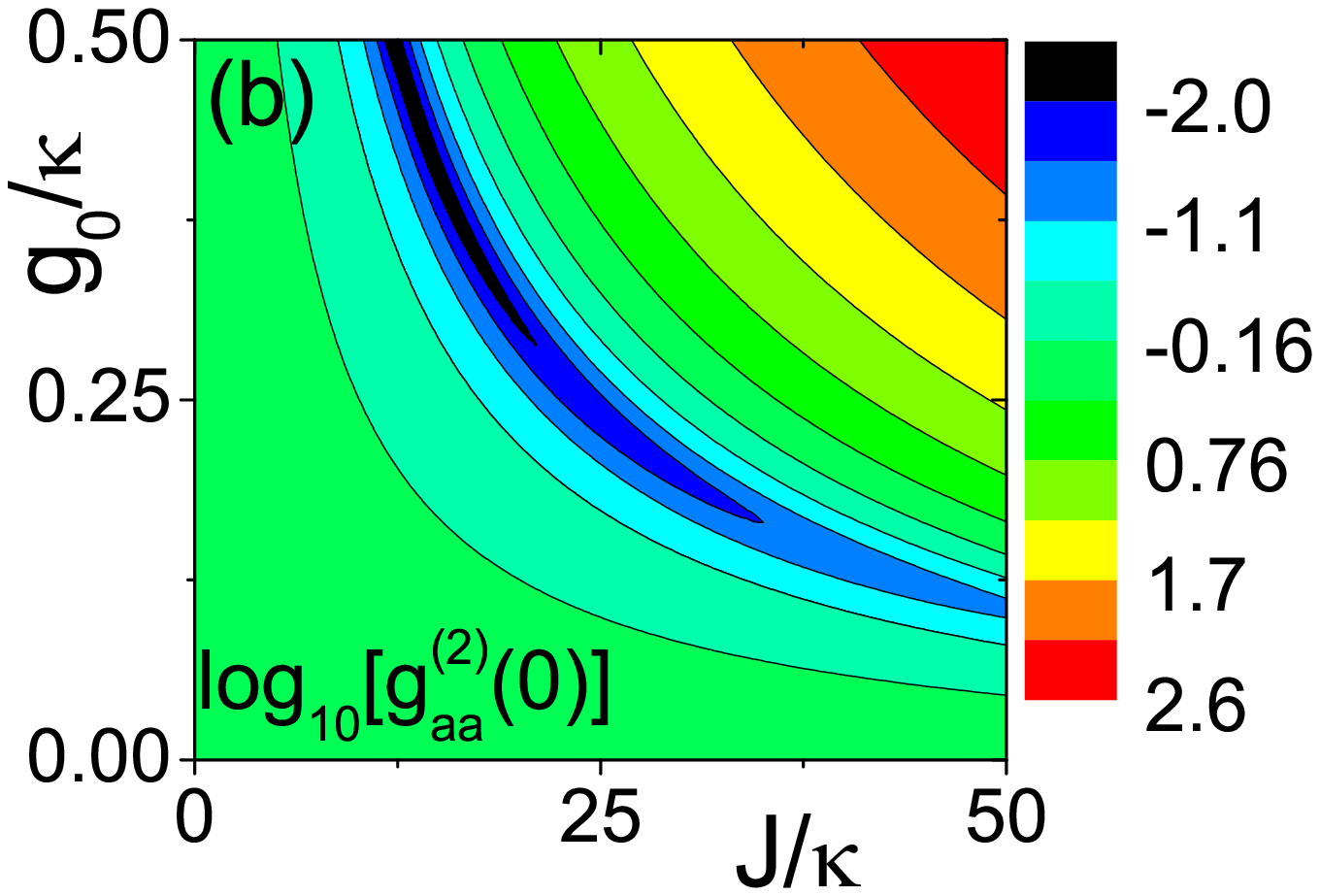}
\includegraphics[bb=10 5 380 280, width=4 cm, clip]{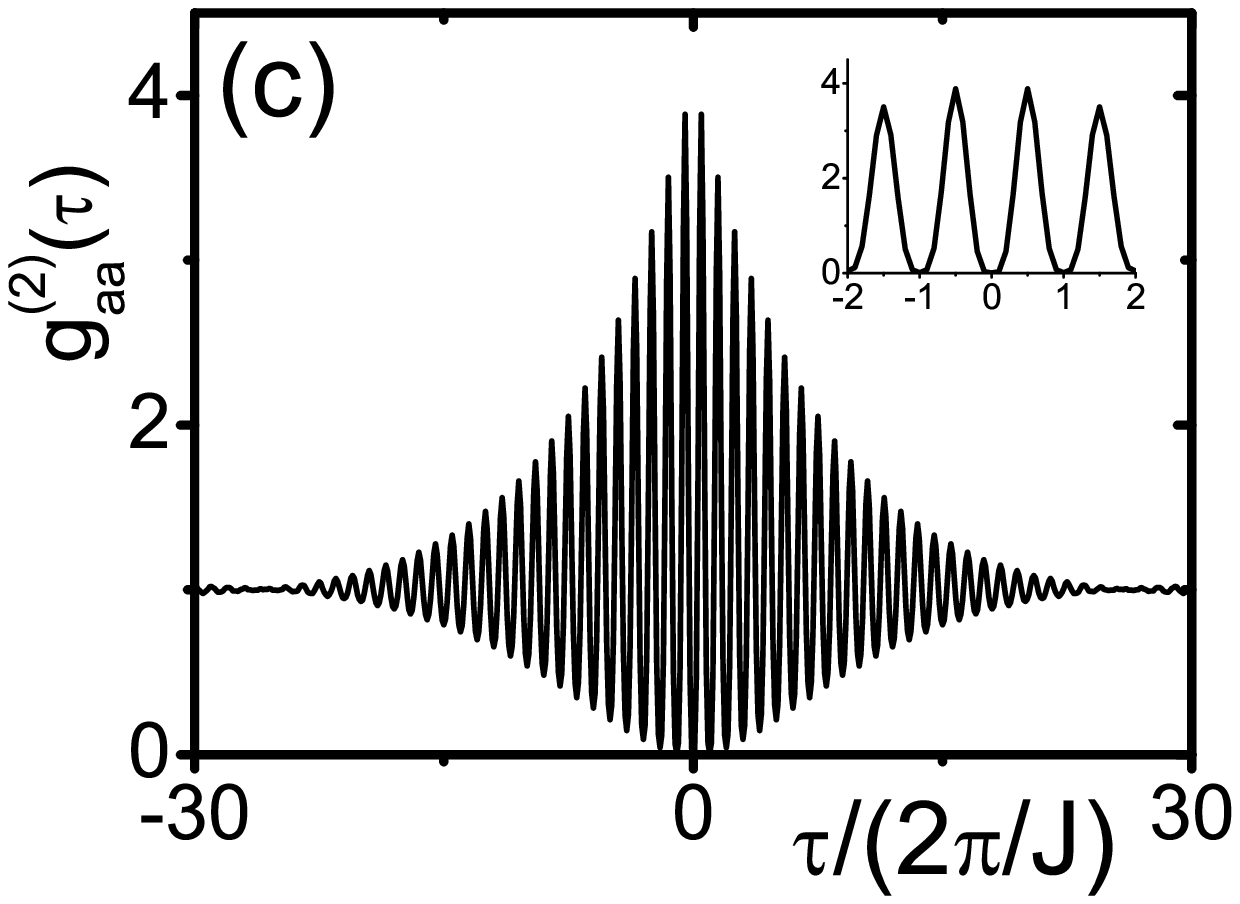}
\includegraphics[bb=10 5 380 280, width=4 cm, clip]{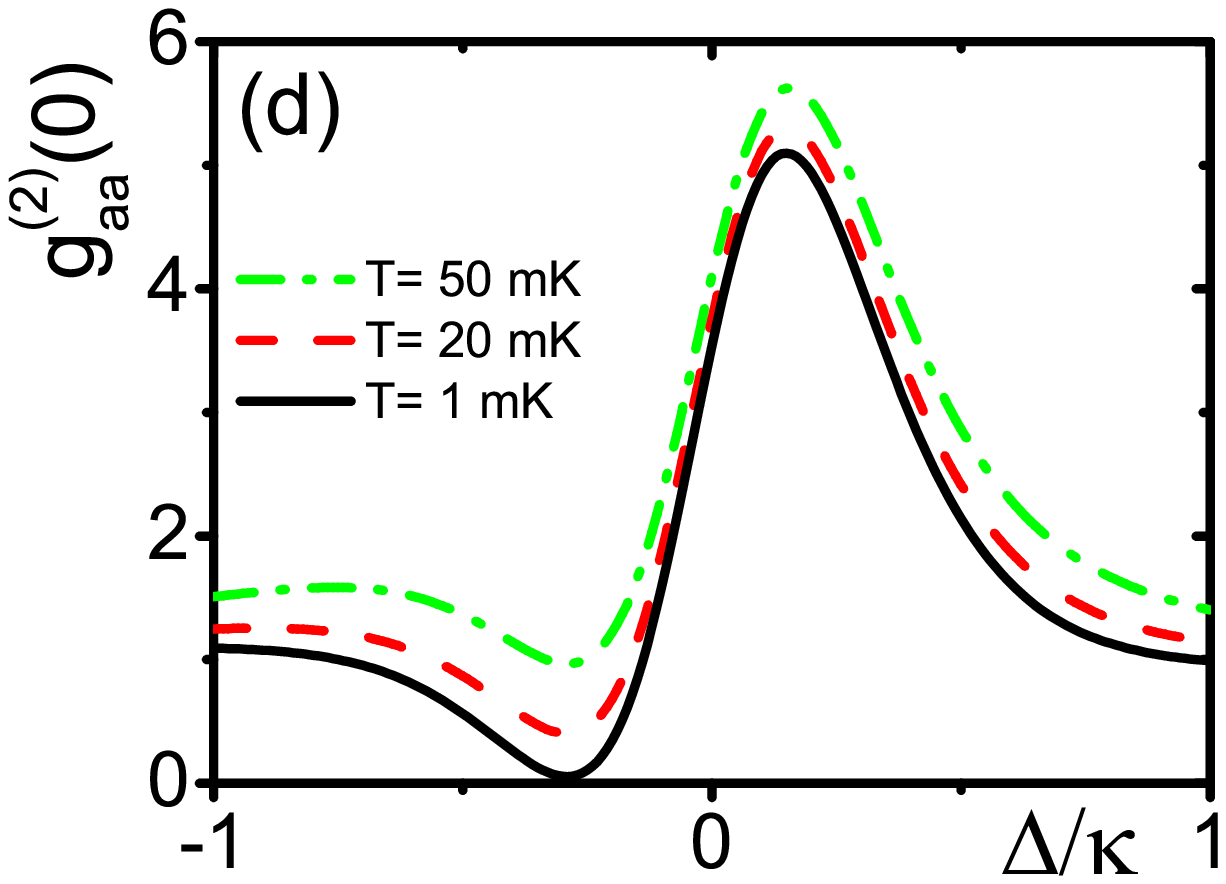}
\caption{$g_{aa}^{\left( 2\right)} \left( 0\right)$ given by Eq.~(\ref{eq:34}), plotted as functions of $\Delta/\kappa$ and $g_{0}/\kappa$ for $J=30 \kappa$ in (a), and as functions of both $g_{0}/\kappa$ and $J/\kappa$ for $\Delta=-0.29 \kappa$ in (b). (c) The time evolution of $g_{aa}^{\left( 2\right)} \left( \tau \right)$ for $J=30 \kappa$, $g_{0}=0.2 \kappa$, $\Delta=-0.29 \kappa$, and in (a-c) $T=1$ mK. (d) $g_{aa}^{\left( 2\right)} \left( 0 \right)$ versus $\Delta/\kappa$ for temperature $T=$ ($1$ mK, $10$ mK, $50$ mK) at $J=30 \kappa$ and $g_{0}=0.2 \kappa$. The other parameters are $\varepsilon_{c}=10^{-2} \kappa$, $\omega_{m}=100 \kappa$, $\omega_{m}/\gamma_{m}=10^{4}$, and $\kappa/2\pi=1$ MHz.} \label{fig2}
\end{figure}

The standard optomechanical system consists of a cavity with an oscillating
mirror at one end~\cite{Rabl, Nunnenkamp}, and as shown in Ref.~\cite{Rabl}, the strong optomechanical coupling ($g_{0} > \kappa$) is the necessary condition for obtaining efficient antibunching photons in the standard optomechanical system. In this section we will show that the cavity can exhibit strong photon antibunching if it is coupled to an optomechanical system even when optomechanical interaction in the system is weak.

Now, let us focus on the statistical properties of light field in
cavity $A$. From now on, we assume that the decay rates
$\kappa_{a}=\kappa_{b}=\kappa$, the
detunings $\Delta_{a}=\Delta _{b}-g_{0}^{2}/\omega _{m}=\Delta $, and
normalize all the parameters to $\kappa$. The equal-time
second-order correlation function of the photons inside cavity $A$,
$g_{aa}^{\left( 2\right) }\left( 0\right)$, is given as functions of $\Delta/\kappa$ and $g_{0}/\kappa$  in
Fig.~\ref{fig2} (a) for  $J=30 \kappa$. We can see that there is an optimal point for the field in cavity $A$ showing strong antibunching at $\Delta=-0.29 \kappa$ and $g_{0}=0.2 \kappa$
for $J=30 \kappa$. The result shows that the photons in cavity $A$ can exhibit
strong antibunching when it is coupled to an optomechanical system
under weak coupling condition ($g_{0} < \kappa$) in the resolved sideband regime ($\omega _{m} \gg \kappa$). As given in Section II, the strong antibunching comes from the cooperation between the weak
optomechanical interaction and the destructive interference between
different paths for two-photon excitation as shown in the reduced diagram in Fig.\ref{fig1}(b).

Two-dimensional plot of the equal-time second-order correlation
function of cavity $A$, $g_{aa}^{\left( 2\right) }\left( 0\right)$,
as functions of both $g_{0}/\kappa$ and $J/\kappa$ is shown in Fig.~\ref{fig2} (b) for
$\Delta=-0.29 \kappa$. With the increasing of $J$, the value of
$g_{0}$ for getting the strong antibunching descends gradually. This implies that the linear coupling between the cavity and the optomechanical system can be used to
lower the strength of the optomechanical interaction that is
required to achieve strong antibunching.

In order to understand the optimal conditions for the strong antibunching, we will find the optimal parameters for the system in the steady-state. As $\Delta_{a}=\Delta _{b}-g_{0}^{2}/\omega _{m}=\Delta $, $J[a^{\dag }b e^{\frac{g_{0}}{\omega _{m}}\left( c^{\dag}-c\right)}+\rm{H.c.}]$ approximately equals $J(a^{\dag }b + \rm{H.c.})$ in the conditions that $\Delta \ll \omega _{m}$, $g_{0}/\omega _{m}\ll 1$ and $J < \omega _{m}/2$. Then in the frame rotating with frequency $\omega_{c}$, the effective Hamiltonian (Eq.~(\ref{eq:2})) can be written approximately as~\cite{ZRGong}
\begin{eqnarray} \label{eq:41}
\widetilde{H}_{\text{eff}} &\approx &\hbar \Delta a^{\dag }a+\hbar \Delta
b^{\dag }b- \hbar \frac{g_{0}^{2}}{\omega _{m}} b^{\dag }b^{\dag }bb+\hbar \omega _{m}c^{\dag }c  \notag \\
&&+\hbar J\left( ab^{\dag }+a^{\dag }b\right) +i\hbar \varepsilon _{c}\left(
a^{\dag }-a\right)
\end{eqnarray}
For the phonon states are decoupled from the photon states, the state of the system can be written as
$\left\vert \psi \right\rangle =\left\vert \varphi \right\rangle
\left\vert \phi\right\rangle _{m}$, where $\left\vert \varphi \right\rangle $ is the photon state, and $\left\vert \phi\right\rangle _{m}$ is the phonon state. Under the weak pumping conditions, using the ansatz:
\begin{eqnarray} \label{eq:42}
\left\vert \varphi \right\rangle  &=&C_{00}\left\vert 0,0\right\rangle
+C_{10}\left\vert 1,0\right\rangle +C_{01}\left\vert 0,1\right\rangle
\notag \\
&&+C_{20}\left\vert 2,0\right\rangle +C_{11}\left\vert 1,1\right\rangle
+C_{02}\left\vert 0,2\right\rangle,
\end{eqnarray}%
and $C_{00}\gg C_{10},C_{01}\gg C_{20},C_{11},C_{02}$, we can get the optimal conditions for $C_{20}=0$
as given in Ref.~\cite{Bamba} as follow
\begin{eqnarray}
\Delta _{\text{opt}}&=&-\frac{1}{2}\sqrt{\sqrt{9J^{4}+8\kappa ^{2}J^{2}}%
-3J^{2}-\kappa ^{2}} \label{eq:43}\\
g_{0,\text{opt}}&=&\sqrt{-\frac{\omega _{m}\Delta \left( 5\kappa ^{2}+4\Delta
^{2}\right) }{2\left( 2J^{2}-\kappa ^{2}\right) }} \label{eq:44}
\end{eqnarray}%
For the two optical cavities in the strong coupling condition $J\gg \kappa $,
the optimal conditions are simplified as
\begin{eqnarray}
\frac{\Delta _{\text{opt}}}{\kappa} & \approx & -\frac{1}{2\sqrt{3}}\approx-0.29, \label{eq:45}\\
\frac{g_{0,\text{opt}}}{\kappa} & \approx & \sqrt{\frac{2}{3\sqrt{3}}} \sqrt{\frac{\omega _{m}}{J}} \sqrt{\frac{\kappa}{J}}, \label{eq:46}
\end{eqnarray}
which perfectly agree with the results shown in Fig.~\ref{fig2} (a, b). The terms on the right of Eq.~(\ref{eq:46}) are a
factor, a ratio greater than $\sqrt{2}$, and the square root of the cavity decay rate to the
coupling between the two cavities. It can be seen clearly that in the resolved sideband limit, the optomechanical coupling constant can be pushed below the single photon strong coupling limit provided that the coupling between the two
cavities is much larger than the optical linewidth of each cavity, but still less than half of the mechanical
frequency.

The time evolution of the second-order correlation
function of cavity $A$, $g_{aa}^{\left( 2\right) }\left( \tau\right)$, is shown in Fig.~\ref{fig2} (c). As reported in Refs.~\cite{Liew,Bamba}, $g_{aa}^{\left( 2\right) }\left( \tau\right)$ oscillates with the period $2\pi/J$. This oscillation comes from the probability oscillation between the photon states $\left\vert 1,0\right\rangle$ and $\left\vert 0,1\right\rangle$. Besides, the timescale of the antibunching is about $2\pi/\kappa$, which is the lifetime of the photon states.

The second order correlation functions for different temperatures
are shown in Fig.~\ref{fig2} (d). From the figure we can see that
the increase of the temperature will suppress the exhibition of
antibunching effect, because the phonons in the environment may disturb the quantum statistics of the system. In order to get strong antibunching effect, keeping the mechanical resonator in the optomechanical system in low temperature environment is one of the necessary conditions.

\begin{figure}
\includegraphics[bb=40 5 410 280, width=4 cm, clip]{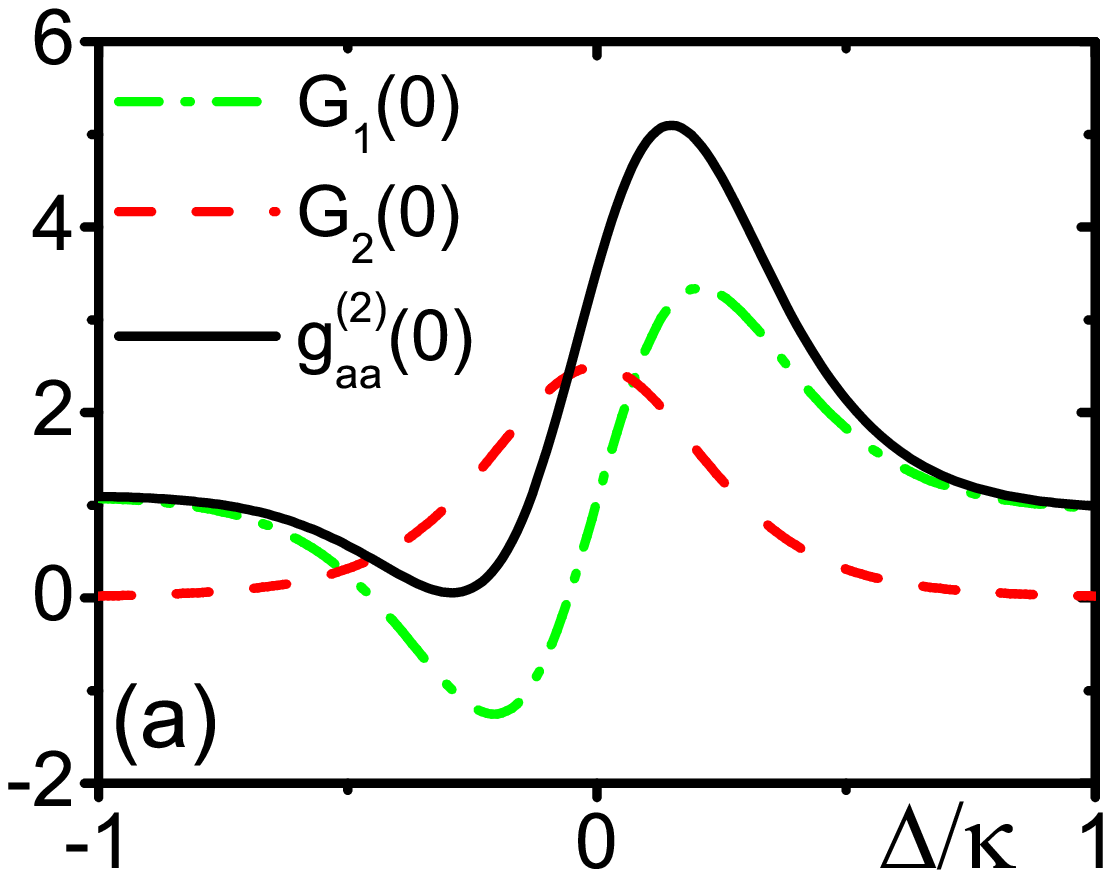}
\includegraphics[bb=40 5 410 280, width=4 cm, clip]{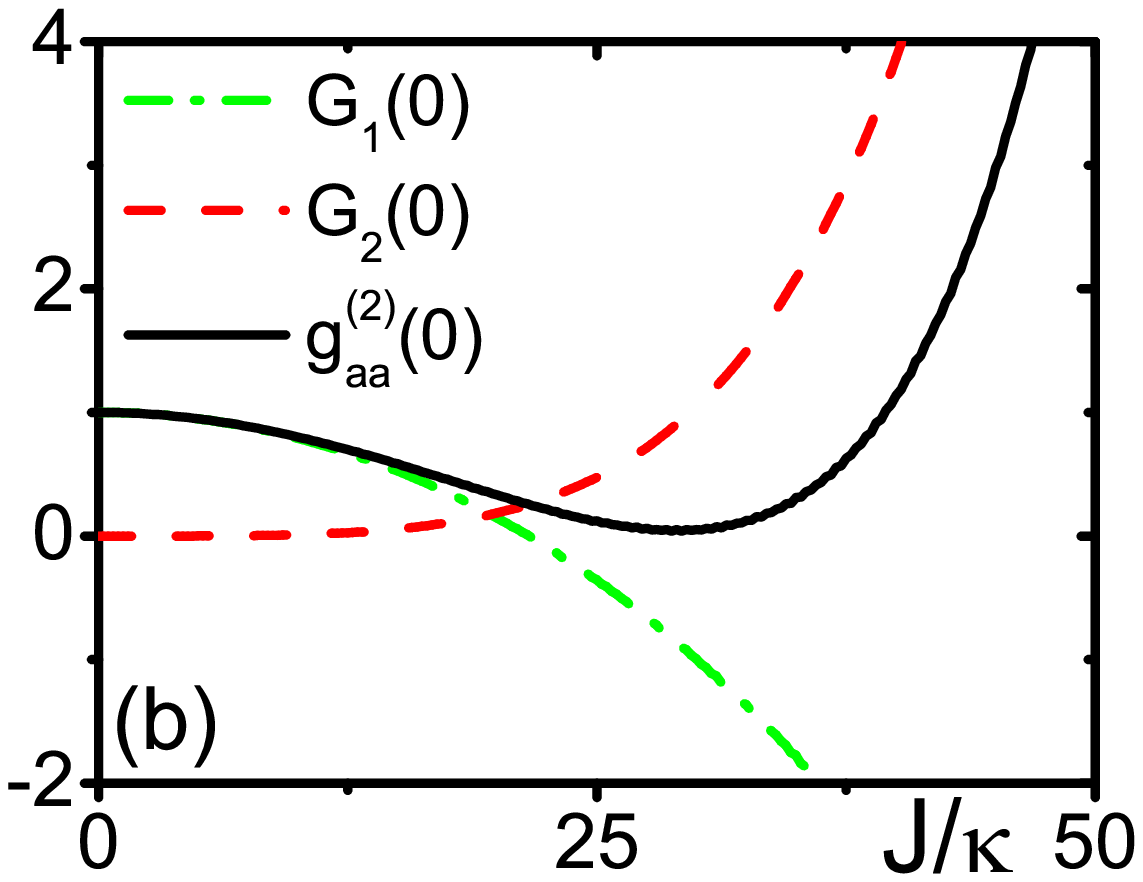}
\includegraphics[bb=40 5 410 280, width=4 cm, clip]{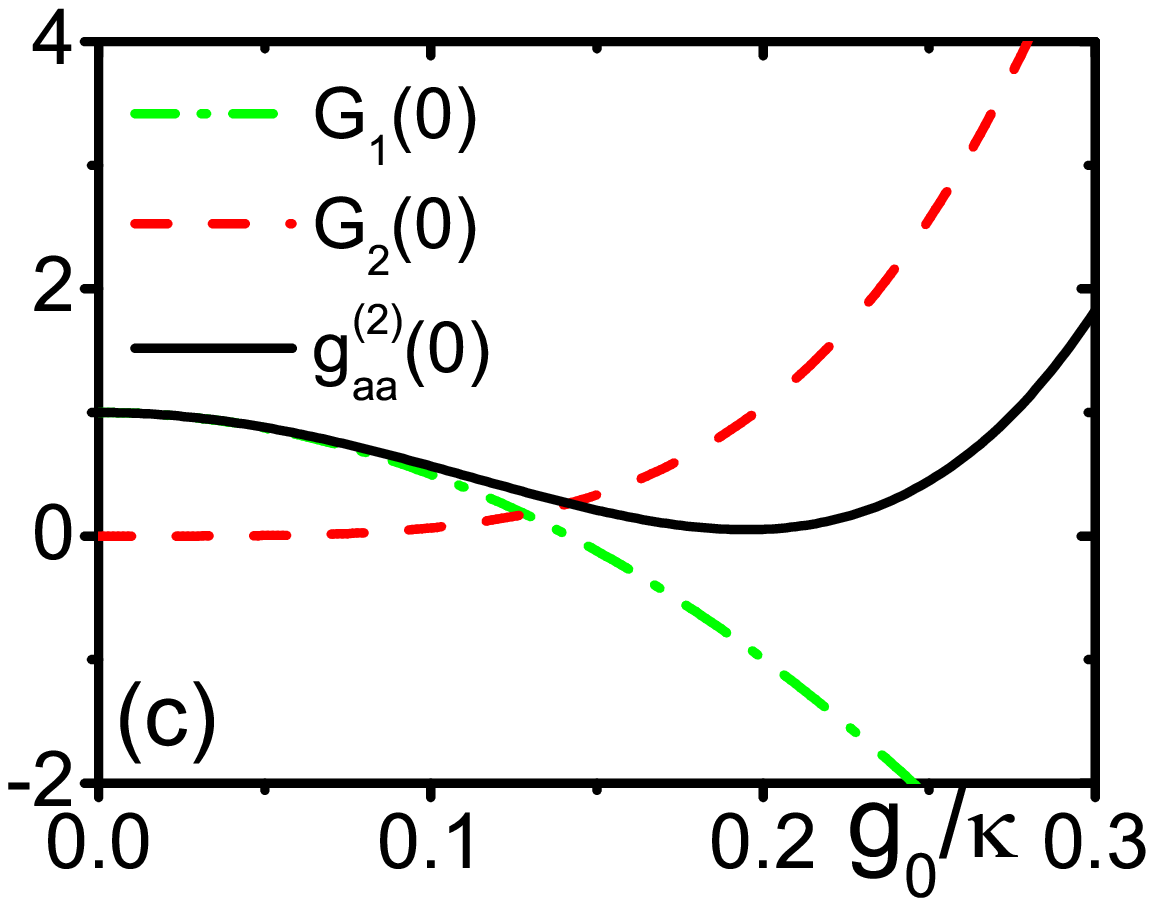}
\caption{$G_{1}(0)$ (dashed-dot), $G_{2}(0)$ (dashed) and $g_{aa}^{\left( 2\right)} \left( 0\right)$ (solid) plotted as
functions of $\Delta/\kappa$ for $J=30 \kappa$ and  $g_{0}=0.2 \kappa$ in (a), as
functions of $J/\kappa$ for $\Delta=-0.29 \kappa$ and  $g_{0}=0.2 \kappa$ in (b), and as
functions of $g_{0}/\kappa$ for $\Delta=-0.29 \kappa$ and  $J=30 \kappa$ in (c). The other
parameters are the same as in Fig.~\ref{fig2}.} \label{fig3}
\end{figure}

In addition, we show $G_{1}(0)$, $G_{2}(0)$ and $g_{aa}^{\left( 2\right)} \left( 0\right)$ as functions of $\Delta/\kappa$ in Fig.~\ref{fig3} (a), as functions of $J/\kappa$ in (b) and as functions of $g_{0}/\kappa$ in (c). In Ref.~\cite{Drummond}, the second-order correlation function is approximately replaced by $G_{1}(0)$ with $G_{2}(0)$ dropped, in the condition that $\left\vert \alpha \right\vert ^{2} \gg \left\langle \delta a^{\dag }\left( t\right) \delta a\left( t' \right) \right\rangle$. From Fig.~\ref{fig3}, we can see that if the driving field is weak, $G_{2}(0)$ plays an important role and should be considered here, otherwise $g_{aa}^{\left( 2\right)} \left( 0\right)$ becomes negative in some condition. As the noises in the system are Gaussian~\cite{Ford1965,Ford1988}, the four-operator correlation are equal to the sum of products of pair correlation functions as shown in Eq.~(\ref{eq:36}).

\section{Numerical solution by Master equation}

We have obtained the analytical expression for the second order correlation function by solving the Langevin equations, but we also have done many approximations, such as semiclassical approximation, factorization assumption, and ignore the high order terms of small fluctuations. In this part, we calculate the second-order correlation function by
numerically solving the master equation of the density matrix, and
compare the results to the predictions given by analytical solution
derived above. The master equation of the coupled system is given
as~\cite{Carmichael}
\begin{eqnarray} \label{eq:47}
\frac{d \rho }{d t} &=&\frac{1}{i\hbar }\left[ \widetilde{H},\rho \right] +%
\frac{\kappa _{a}}{2}\left( 2a\rho a^{\dag }-a^{\dag }a\rho -\rho
a^{\dag }a\right)  \nonumber\\
&&+\frac{\kappa _{b}}{2}\left( 2b\rho b^{\dag }-b^{\dag }b\rho -\rho
b^{\dag }b\right)  \nonumber\\
&&+\frac{\gamma _{m}}{2}\left( 2c\rho c^{\dag }-c^{\dag }c\rho -\rho
c^{\dag }c\right)  \nonumber\\
&&+\gamma _{m} \bar{n}_{m}\left( c\rho c^{\dag }+c^{\dag }\rho
c-c^{\dag }c\rho -\rho cc^{\dag }\right),
\end{eqnarray}%
where $\bar{n}_{m}$ is the mean thermal phonon number of the moving
mirror given by the Bose-Einstein statistics
$\bar{n}_{m}=[\exp(\hbar \omega_{m}/k_{B}T)-1]^{-1}$, $k_{B}$ is the
Boltzmann constant and $T$ is the effective temperature of the
moving mirror. The master equation can be solved in the basis of the
photon and phonon number states $|n_{a},n_{b},n_{m}\rangle$, and
$\rho$ can be written as density matrix
\begin{equation} \label{eq:48}
\rho =\rho _{n_{a},n_{b},n_{m};n_{a}^{\prime },n_{b}^{\prime
},n_{m}^{\prime }}(t)|n_{a},n_{b},n_{m}\rangle \langle n_{a}^{\prime
},n_{b}^{\prime },n_{m}^{\prime }|.
\end{equation}
If the elements of the steady state density matrix, $\rho
_{n_{a},n_{b},n_{m};n_{a}^{\prime },n_{b}^{\prime },n_{m}^{\prime
}}$, are given, the equal-time second-order correlation function can
be easily calculated by
\begin{eqnarray}
g_{aa}^{\left( 2\right) }\left( 0\right) &=& \frac{{\rm Tr} [\rho
a^{\dag
2}a^{2}] }{ [{\rm Tr} (\rho a^{\dag }a)]^2}, \label{eq:49} \\
g_{bb}^{\left( 2\right) }\left( 0\right) &=& \frac{{\rm Tr} [\rho
b^{\dag
2}b^{2}] }{ [{\rm Tr} (\rho b^{\dag }b)]^2}, \label{eq:50} \\
g_{ab}^{\left( 2\right) }\left( 0\right) &=& \frac{{\rm Tr} [\rho
a^{\dag}b^{\dag}ba] }{[{\rm Tr} (\rho a^{\dag }a)][{\rm Tr}
(\rho b^{\dag }b)]}, \label{eq:51}
\end{eqnarray}%
where $g_{ab}^{\left( 2\right) }\left( 0\right)$ is the cross
correlation between the photons in cavity $A$ and $B$.

\begin{figure}
\includegraphics[bb=10 5 380 280, width=4 cm, clip]{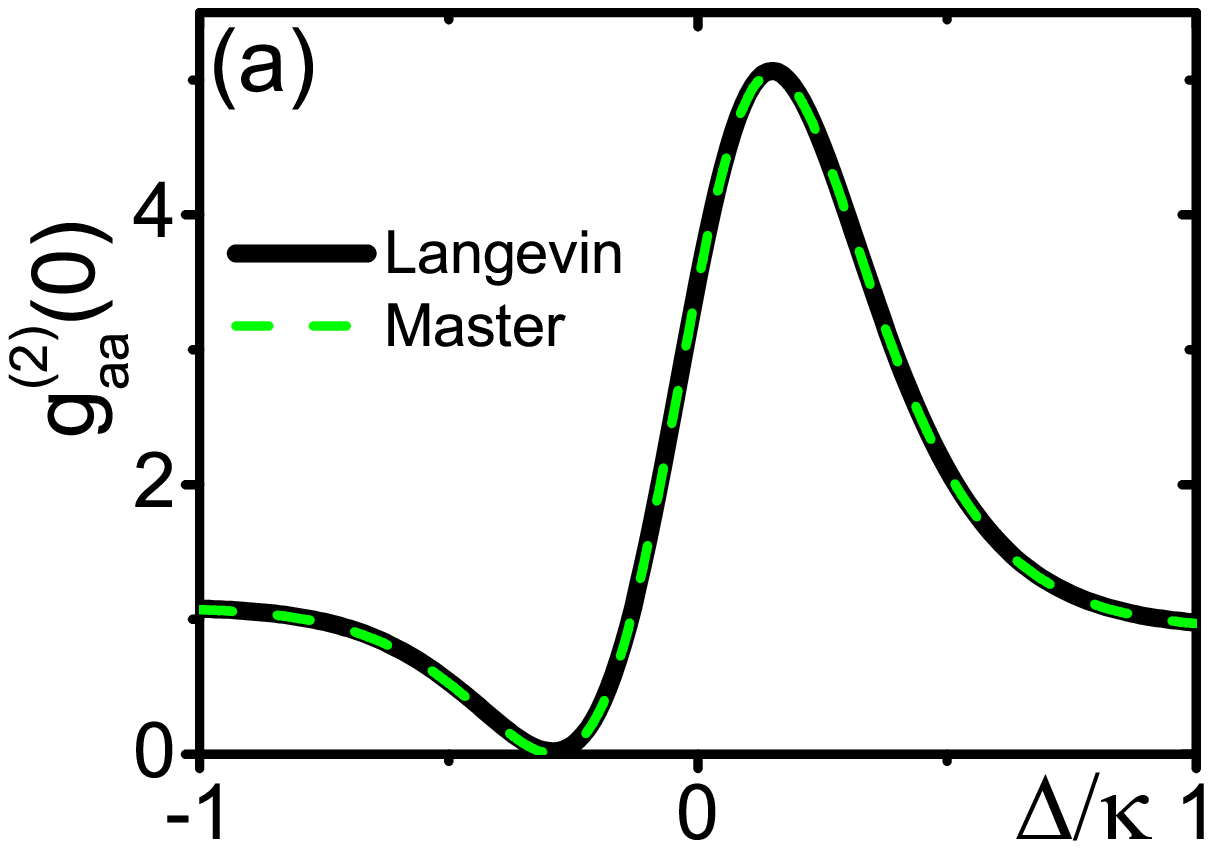}
\includegraphics[bb=10 5 380 280, width=4 cm, clip]{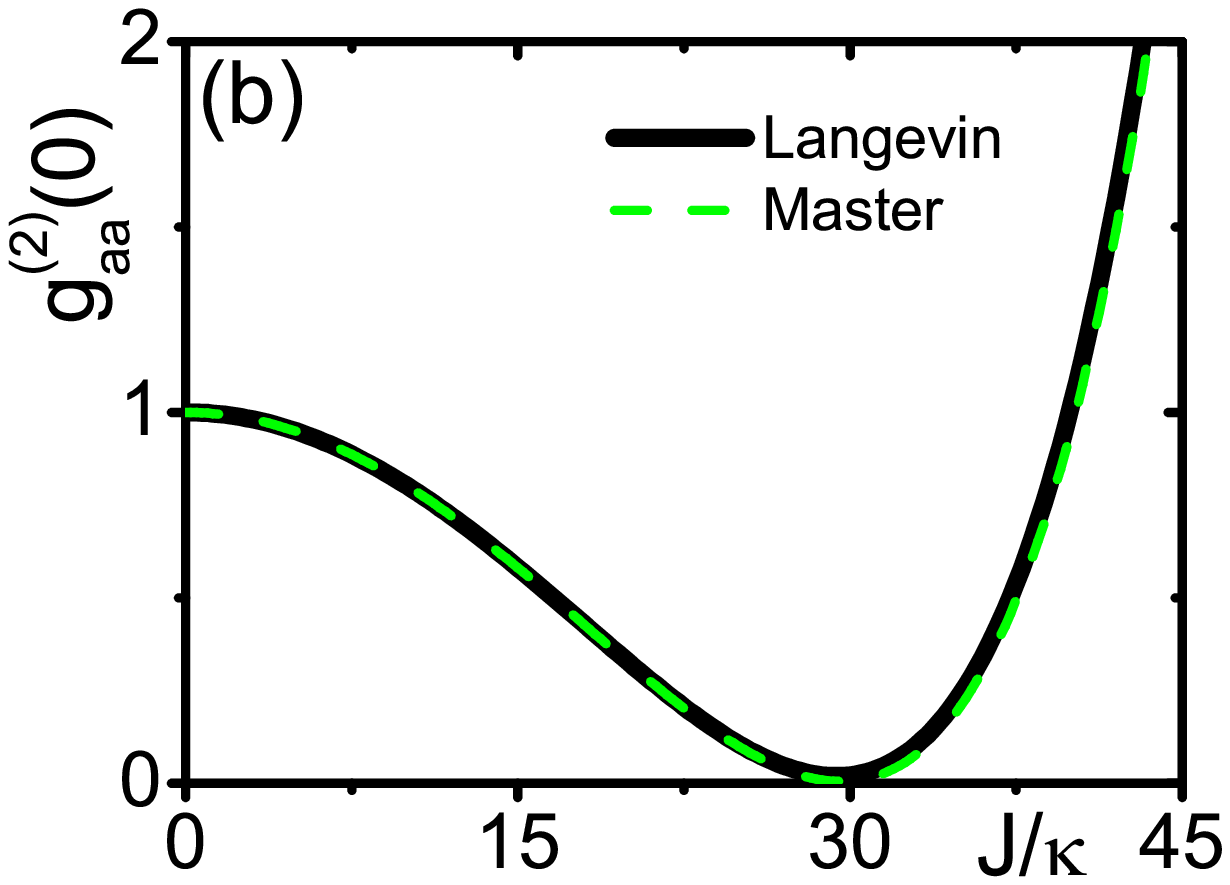}
\includegraphics[bb=10 5 380 280, width=4 cm, clip]{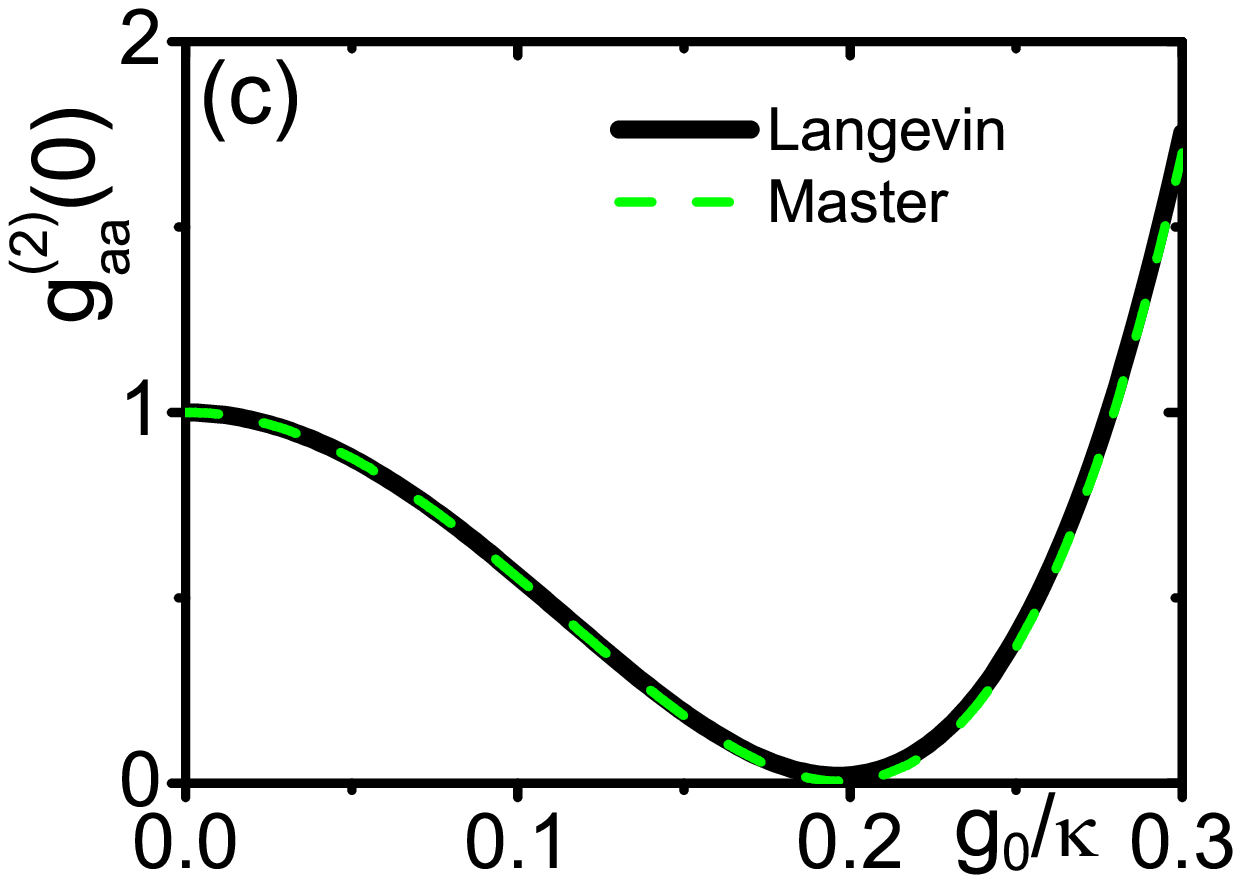}
\includegraphics[bb=10 5 380 280, width=4 cm, clip]{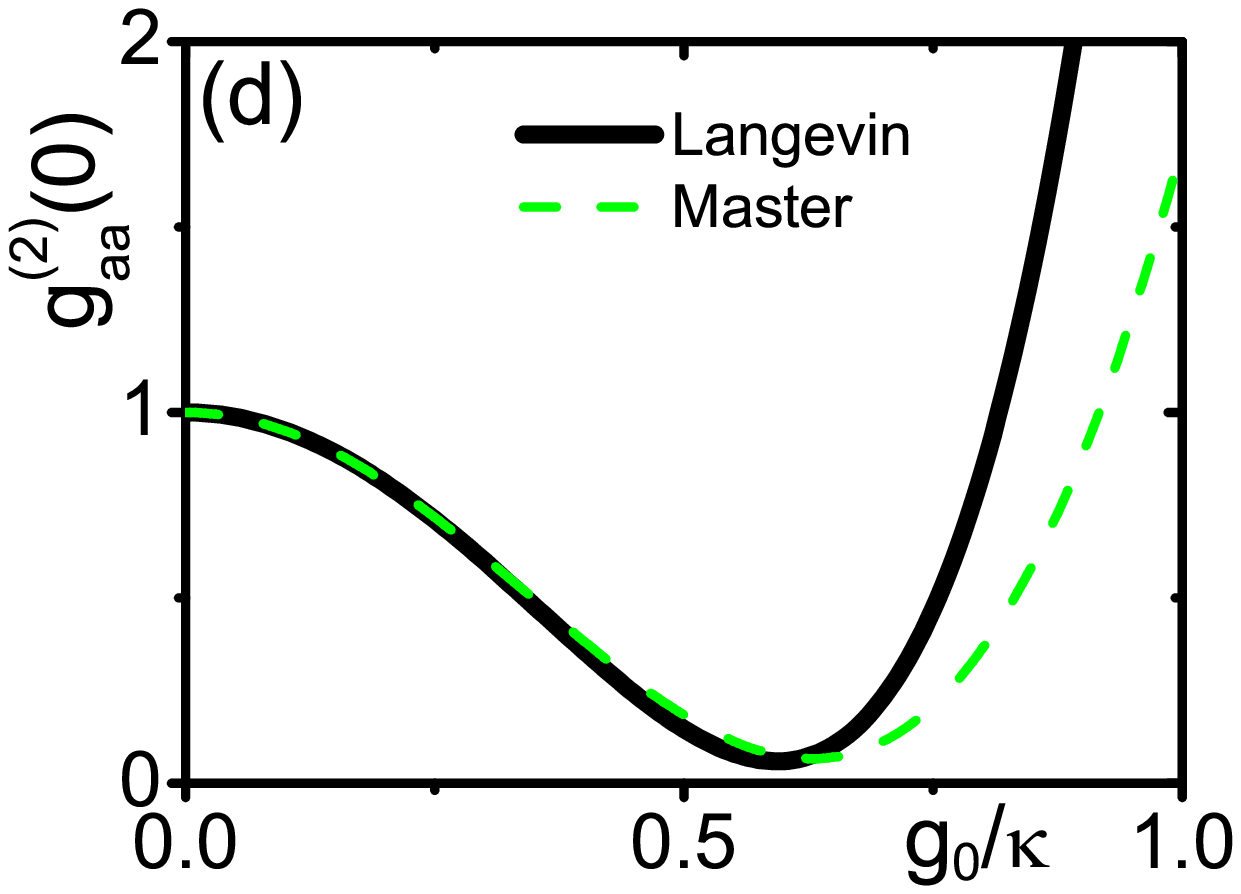}
\caption{$g_{aa}^{\left( 2\right) }\left( 0\right)$ calculated by
the quantum Langevin equations (solid) and master equation (dashed) plotted as
functions of (a) $\Delta/\kappa$, (b) $J/\kappa$, and (c, d) $g_{0}/\kappa$. The parameters in (a)-(c) are the same as in Fig.~\ref{fig3}(a)-(c); the parameters in (d) are $J=3 \kappa$, $\omega_{m}=10 \kappa$, $\omega_{m}/\gamma_{m}=10^{3}$ and $T=0.1$ mK.} \label{fig4}
\end{figure}

For comparison, the second order correlation functions calculated by
the master equation and quantum Langevin equations are shown in the
same figure as functions of $\Delta/\kappa$ in Fig.~\ref{fig4} (a), as functions of $J/\kappa$ in (b) and  as functions of $g_{0}/\kappa$ in (c) and (d). From Fig.~\ref{fig4} (a)-(c), we can see that the results obtained by the two methods match quantitatively. As shown in Fig.~\ref{fig4}
(d), for $g_{0}<0.65 \kappa$, the predictions by the two methods agree with each other. But with further increasing of $g_{0}$, the
difference between them becomes significant gradually, and linearized quantum
Langevin equations method can only describe this qualitatively.

\begin{figure}
\includegraphics[bb=0 0 390 280, width=4 cm, clip]{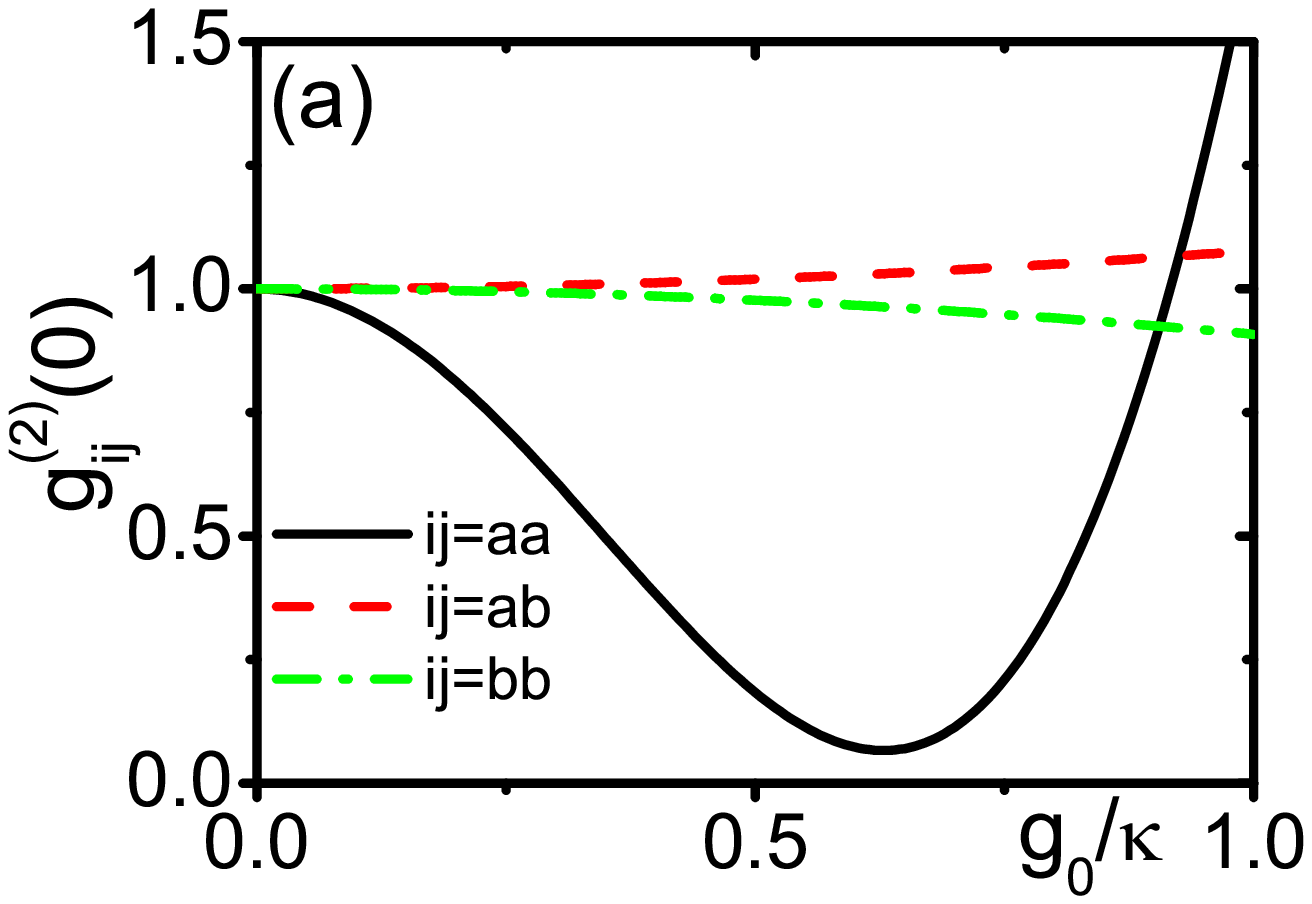}
\includegraphics[bb=10 5 390 270, width=4 cm, clip]{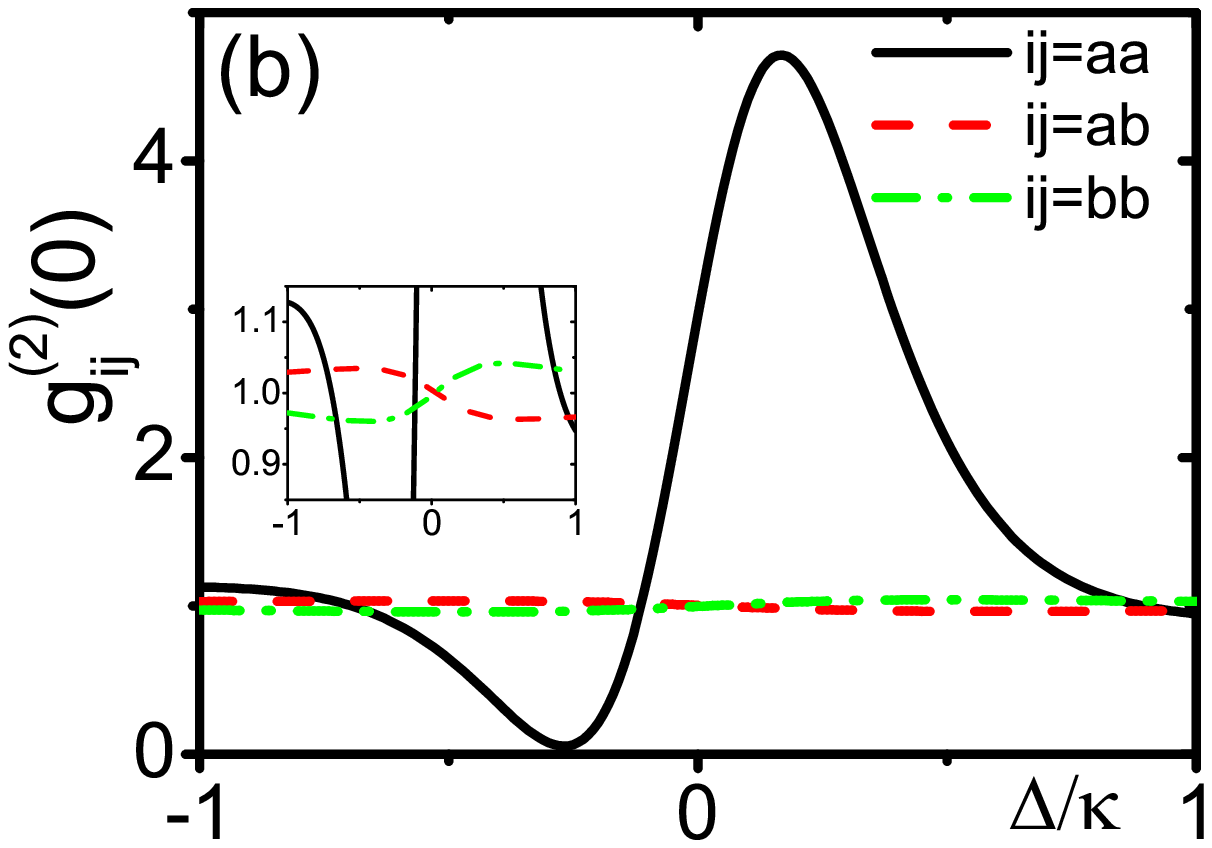}
\caption{(a) $g_{ij}^{\left( 2\right) }\left( 0\right)$ plotted as functions of
$g_{0}/\kappa$ at $\Delta=-0.29 \kappa$, $ij=aa$ for cavity $A$ (solid),
$ij=ab$ for the cross correlation between the two cavities (dashed),
and $ij=bb$ for cavity $B$ (dashed-dot). (b) Dependence of
$g_{ij}^{\left( 2\right) }\left( 0\right)$ on $\Delta/\kappa$ for $g_{0}=0.63
\kappa$. The other parameters are the same as in Fig.~\ref{fig4}
(d).} \label{fig5}
\end{figure}

Finally, let us take a look at the statistical properties of
photons in the entire system. The equal-time second order
correlation functions $g_{ij}^{\left( 2\right) }\left( 0\right)$ can
be calculated by using Eqs.~(\ref{eq:49}-\ref{eq:51}) and the
results are shown in Fig.~\ref{fig5}. From Fig.~\ref{fig5} (a) we
can see that, under weak optomechanical interaction condition, there is strong antibunching in cavity $A$ around
$g_{0}=0.6 \kappa$, while weak antibunching in cavity $B$ and bunching
for the photons between the two cavities.

Dependence of $g_{ij}^{\left( 2\right) }\left( 0\right)$ on $\Delta/\kappa$
is drawn in Fig.~\ref{fig5} (b). Fig.~\ref{fig5} (b) shows us two
interesting phenomena as $\Delta$ is in different domains: For
$\Delta<-0.1 \kappa$, there is strong antibunching in cavity $A$
and weak antibunching in cavity $B$, while the cross correlation
between the modes in the two cavities exhibits bunching. On the
contrary, when $\Delta>0.05 \kappa$, there is bunching in cavity $A$
and cavity $B$, while the cross correlation between the modes in the
two cavities exhibits weak antibunching $g^{(2)}_{ab}(0)<1$. Under
weakly driven condition~\cite{xu-liu}, $g^{(2)}_{aa}(0)<1$ and
$g^{(2)}_{bb}(0)<1$ indicate that there is no more than one photon
in each cavity, and $g^{(2)}_{ab}(0)>1$ shows that there is big
chance that each cavity has one photon simultaneously.
$g^{(2)}_{aa}(0)>1$ and $g^{(2)}_{bb}(0)>1$ indicate that there is
big chance for more than one photon present in each cavity, while
$g^{(2)}_{ab}(0)<1$ shows that the probability that each cavity has
one photon simultaneously is low. In other words, when the system is
driven weakly, and there are two photons in the coupled system, if
$\Delta<-0.1 \kappa$, they are likely to be in the state that each
cavity has one photon simultaneously, and if $\Delta>0.05 \kappa$,
they prefer to stay in one of the cavities together at the same
time. The similar phenomena have been reported in the system with Kerr nonlinearity~\cite{Bamba}.

\section{Conclusions}

We have studied the photon statistics of a cavity linearly coupled
to an optomechanical system. Due to destructive quantum interference
effect between different paths for two-photon excitation, the cavity
can exhibit strong photon antibunching with weak optomechanical
interaction in the optomechanical system. Both analytical and
numerical methods are employed to figure out our results. The results
bring hope to us of observing photon blockade effect with current
experimental parameters of optomechanics.

We gratefully thank Professor Yu-xi Liu for comments and Dr Yan-Jun Zhao for valuable discussions.

\bibliographystyle{apsrev}
\bibliography{ref}

\end{document}